\documentclass[10pt,twocolumn,letterpaper,anonymous=true]{article}

\usepackage{tabu}
\usepackage{booktabs}
\usepackage{cvpr}
\usepackage{times}
\usepackage{epsfig}
\usepackage{graphicx}
\usepackage{amsmath}
\usepackage{amssymb}
\usepackage{subcaption} 

\usepackage[breaklinks=true,bookmarks=false]{hyperref}

\cvprfinalcopy 

\def\cvprPaperID{****} 

\begin{document}

\title{Your Eyes Say You're Lying: An Eye Movement Pattern Analysis for 
Face Familiarity and Deceptive Cognition}

\author{Jiaxu Zuo\\
Australian National University\\
Canberra, Australia\\
{\tt\small jiaxu.zuo@anu.edu.au}
\and
Tom Gedeon\\
Australian National University\\
Canberra, Australia\\
{\tt\small tom@cs.anu.edu.au}
\and
Zhenyue Qin\\
Australian National University\\
Canberra, Australia\\
{\tt\small zhenyue.qin@anu.edu.au}
}

\maketitle

\begin{abstract}
   Eye movement patterns reflect human latent internal cognitive activities. We aim to discover eye movement patterns during face recognition under different cognitions of information concealing. These cognitions include the degrees of face familiarity and deception or not, namely telling the truth when observing familiar and unfamiliar faces, and deceiving in front of familiar faces. We apply Hidden Markov models with Gaussian emission to generalize regions and trajectories of eye fixation points under the above three conditions. Our results show that both eye movement patterns and eye gaze regions become significantly different during deception compared with truth-telling. We show the feasibility of detecting deception and further cognitive activity classification using eye movement patterns. 
\end{abstract}

\section{Introduction}

During criminal or other forms of investigations for justice, suspects may deceive the investigators by claiming not recognizing a familiar face in order to clear their own suspicion or to protect their fellow co-conspirators. Failure of detecting this delinquent deception can cause a severe threat to the society since it can imprison innocent citizens whilst let the guilty defendant go free~\cite{wu2018deception}. For example, Several days before the terror attack in Brussels, France in 2016, one of the accomplices was arrested and interrogated about his relationship with the terrorist group. He denied any familiarity with the photos of that terrorist group shown to him~\cite{lichfield2018paris}. 

Lie detection, such as to interrogate suspects the familiarity of other people, plays an essential role in maintaining the justice and stability of the society. However, human capability of discerning deception is poor, with an accuracy of slightly being better than chance~\cite{bond2006accuracy}. Thus, people have raised various methodologies aiming to support criminal investigators for fighting malicious tricks. Physiologists have shown that lying will lead to a range of physiological changes of the body~\cite{bhutta2015single}. These changes caused by telling lies facilitate the emergence of physiological approaches~\cite{walczyk2013advancing}. For example, functional Magnetic Reasonance Imaging (fMRI) based methods can present very accurate results~\cite{davatzikos2005classifying}. Nonetheless, these methods contain overmuch noises, heavy expenses, and other drawbacks, which make them infeasible to use in practice~\cite{farah2014functional}. 

Another perspective of discerning deception is through leveraging behaviour cues, which may seem to be negligible to normal people~\cite{o200412}. For example, facial micro-expressions, like eyebrows raising, may reflect that the subject is trying to hide their real emotions~\cite{ekman2009telling}. Nevertheless, it is very laborious to train experts with these behavioural skills due to the significant variability among different subjects~\cite{wu2018deception}. That is, the same micro-emotion may have different cognitive meanings with respect to different persons. 

Unlike fMRI and facial micro-expressions, eye fixation locations and trajectories are relatively easy to capture and collect by ordinary people with the assistance of simple eye trackers. Previous studies illustrate that trajectories of eye gaze movements reflect individuals' underlying cognitive activities~\cite{just1976eye}. For instance, previous research shows that there will be fewer eye fixations during the process of recognizing familiar faces than unfamiliar faces~\cite{hannula2010worth}. 

Another key feature of  eye movements lies on its involuntary property. That is, participants are not able to easily alter their behaviours, irrespective to the undertaken tasks~\cite{ryan2007obligatory}. Due to these advantages of utilizing eye movements, we study the variability of eye movements for deception and frankness given a familiar or unfamiliar face. This variability includes the visiting orders of facial regions and the distributions of eye fixation locations. 

Several recent publications have attempted to investigate the relationship between face familiarity and eye fixations. Millen et al. collected the number of fixations in total and in different facial regions of stating to be unfamiliar under four conditions, namely unfamiliar faces, newly learned faces, famous celebrity faces, and personal familiar faces~\cite{millen2017tracking}. However, they did not include the analysis of the eye movement paths and the facial region of interests in their studies. Lancry-Dayan et al. recently explored in 2018 the possibility of detecting a subject's familiarity with other faces using machine learning techniques by demonstrating him or her four face photos in parallel~\cite{lancry2018you}. They discovered that among the given four faces, subjects turn to fixated more on their familiar faces, followed by a tendency to move eyes away from it~\cite{lancry2018you}. Furthermore, machine learning techniques can successfully leverage this behaviour and showed robust results among different individuals~\cite{lancry2018you}. Nevertheless, they did not explicitly scrutinize the eye movement behaviors on a single individual's face. That is, their study is to investigate the eye movement patterns on four parallel displaying photos instead of one single photo.

In this paper, we aim to investigate the eye movement behaviors during different face recognition tasks under distinct cognitive loads, namely truth telling or lying. Specifically, we explore the eye gaze patterns under three different situations, which are: (a) telling truth on a familiar face; (b) telling truth on an unfamiliar face; (c) lying on a familiar face, where we have ignored the case of ``lying on an unfamiliar face'' due its unreality in the real world. Unlike the previously published work, we focus on the trajectories of eye fixations and their distributions on facial region of interests when undertaking different cognitive tasks. That is, we show that individuals stare at different facial regions when telling the truth or a lie, and the visiting orders of these regions are different. 

Our work can contribute to the future design of automatic deception detection systems. For example, Wu et al. take inputs of multiple modalities, including individual motions, audios and so on, to discern the veracity of expressions from real-life courtroom trial videos~\cite{wu2018deception}. Our work may serve as an additional input channel. Furthermore, our study can also shed lights to the field of visual saliency predictions within computer vision, which focuses on predicting which objects that people will fixate given an image or video~\cite{kruthiventi2016saliency}. Our results indicate that additional channels of people's cognitive activities may further increase the saliency prediction accuracy. Moreover, our research also reveals the possibility of interpret people's cognitive activities from eye movement patterns.

\section{Methods}

\subsection{Fixation Identification}

The raw eye gaze data is in the form of the Cartesian coordinate system, namely its locations are expressed by $(x, y)$ coordinates. It is essential to separating and labelling eye tracking points as fixations or saccades, since improper classification can have dramastic influence on higher-level analyses~\cite{salvucci2000identifying}. 

\begin{figure}
    \centering
    \includegraphics[width=1.0\linewidth]{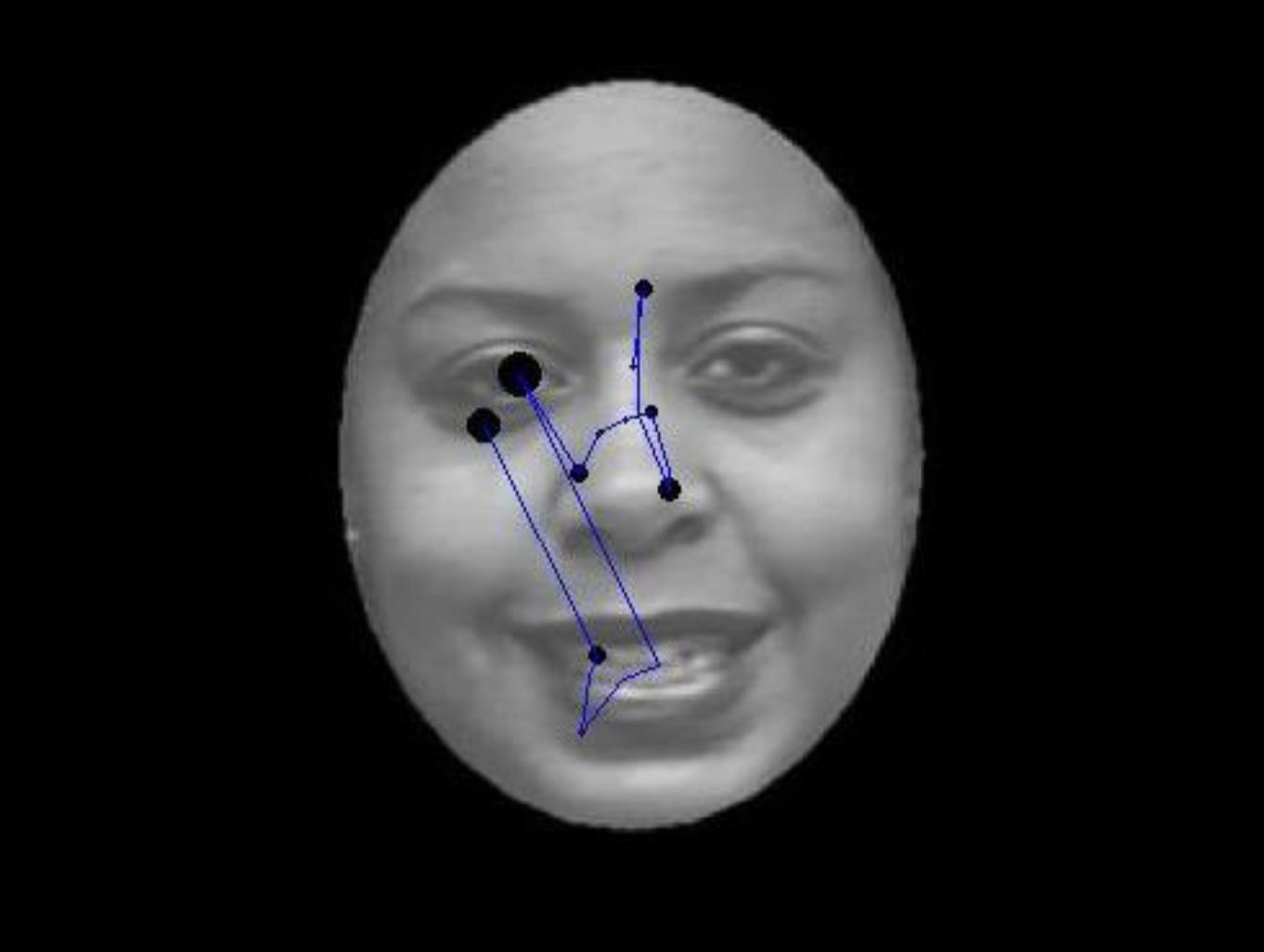}
    \caption{Fixations and the transition paths (i.e. saccades) by using the dispersion threshold algorithm. Black dots are fixation points and blue lines represent eye saccade paths. }
    \label{fig:fixations_transition_paths}
\end{figure}

The dispersion threshold algorithm is considered to be the most robust and accurate approach for identifying fixations and saccades in eye-tracking protocals compared with other approaches such as velocity-based and area-based ones~\cite{salvucci2000identifying}. One essential parameter for this algorithm is called the pixel tolerance, which is a threshold that determines whether to classify a new eye-stare point as a fixation or a saccade. Previous research reported that a normal fixation duration is about 200-250 ms~\cite{miyao1989effects}, and each of our stimulus video lasts for five seconds. Thus, we can deduce that 20 fixation points per video should be an reasonable number, which leads to a setting of the pixel tolerance being 5. Figure \ref{fig:fixations_transition_paths} presents an example of the result using the dispersion threshold algorithm. 

\subsection{Eye Movement Analysis}

\label{subsec:eye_movement_analysis}

We apply probabilistic graphical models, specifically Gaussian Hidden Markov models (HMMs with 2D Gaussian emission distribution), to train the previous collected eye movement data from participants. This technique is broadly utilized to model data generated from Markov processes~\cite{Bishop:2006:PRM:1162264, chuk2014understanding}. That is, the next state of a process only depends on its previous state. For every pair from a current state $S$ to its subsequent state $S'$ in the next time step, or to the observable state in the same time step $O$, there is an associated probability indicating the likelihood of transforming $S$ to $S'$ or $S$ to $O$, which can be formalized as a matrix. Finally, a vector of prior probabilities reveals the different probabilities of being on distinct starting states. We hypothesize to obtain different representative eye movement patterns during face recognition with distinct latent mental activities, namely telling the truth or lying. In this context, states represent different ROIs of a face image, observable states are fixation points, and the transition from the current latent state to the next state is a saccade. 

This methodology requires determining the number of facial regions of interests, i.e., Region of Interests (ROIs) beforehand. We pick the number as two and three. Picking three is since previous studies show that there are three information-rich inner regions of the face (eyes, nose, and mouth), which are particularly essential for person recognition~\cite{o2001familiarisation}. 


In our case, we set the concentration parameters of Dirichlet prior distributions of initial distribution and transition probability to be 0.01. The prior covariance was set as an isotropic covariance matrix with standard deviation of 14 to ensure that the ROI has roughly the same size as the facial features. 

We first trained a Gaussian-HMM using the fixation data from one participant who was viewing images under a certain cognitive condition. For example, the participant has been instructed to tell the truth, followed by observing familiar faces. Fixation sequences during every observation is considered as viewing sequences. Followed by applying the Gaussian-HMM algorithm on multiple observation sequences, we can get a representative Gaussian-HMM with three states. Each state consists of three means and three covariance matrices representing the ROI centers and ranges. 

For every participant, there are four Gaussian-HMMs representing four different eye movement strategies, corresponding two degrees of face familiarity and lying or not. As a result, we have 84 Gaussian-HMMs from 21 participants, and classify them into four classes based on the face familiarity and lying cognitions. Afterwards, we train another Gaussian-HMM for each category in order to find a class representative Gaussian-HMM and its clustering centroids of its ROIs. Eventually, we will have four representative Gaussian-HMMs, each will contain its clustering centroids of ROIs, shapes of ROIs, an intial distribution vector of ROIs, a transition probability matrix, and a Viterbi path.

\section{Experiments}

\subsection{Face Image Preprocessing}

\begin{figure}
    \centering
    \includegraphics[width=1.0\linewidth]{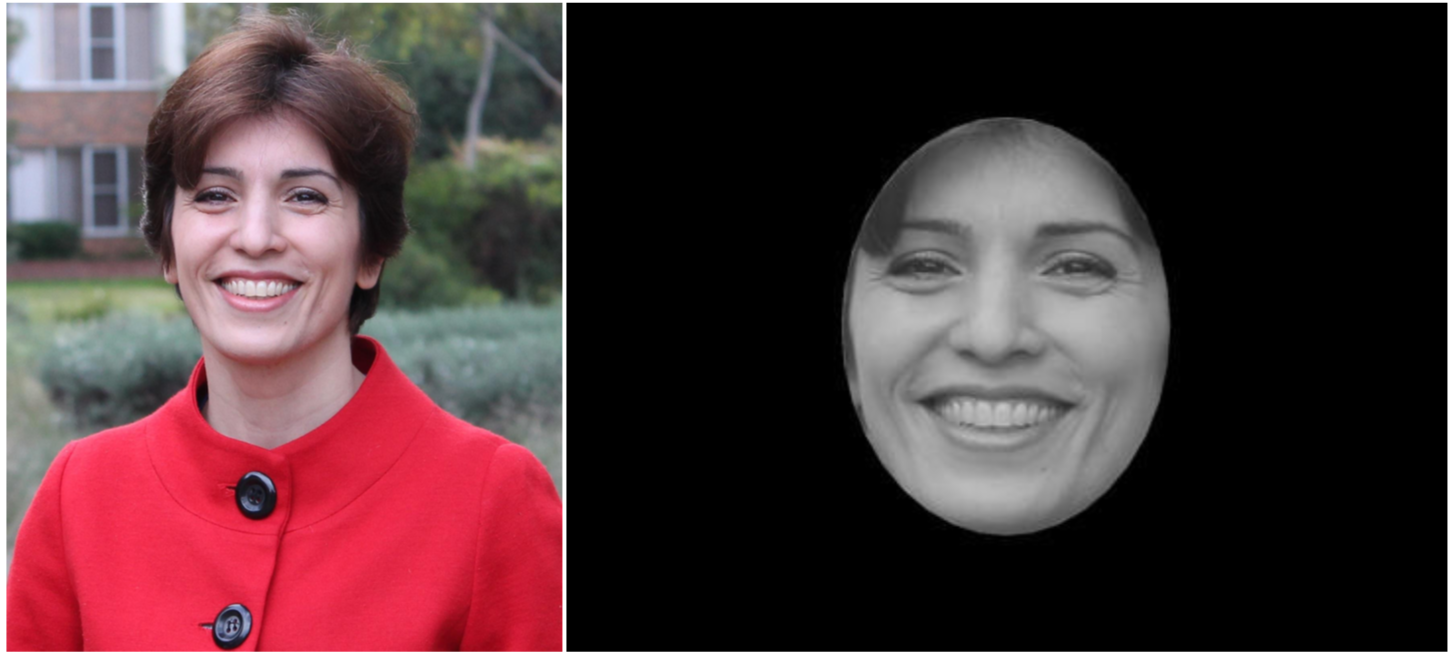}
    \caption{One original face image and its processed result. }
    \label{fig:dean_and_masked}
\end{figure}

The variety of brightness and colors of photos can cause differences in human eye saccades and fixations, as well as a significant latency of the reaction time for viewing the same images~\cite{kalloniatis2007light}. We wish to minimize the noises resulting from this process, known as adaptation in ocular physiology. Therefore, we normalize all the experimental face images to be consistent with the same grey-scale and size of $447\times335$. Furthermore, we apply an oval mask to all the other parts of the images, including hairs, backgrounds and so on, except the main face area. This is to avoid the eye gaze being distracted by the background and objects other than our interested regions, namely human faces. One original face image and its processed result is as Figure \ref{fig:dean_and_masked}. 

\subsection{Face Images and Participants}

We collected a total of 20 different gray-scale frontal-view face images, each will be displayed two times in a random order, one corresponding to telling the truth and the other for lying. That is, one participant will observe 40 images in total during the experiment. Among These faces, 15 out of 20 correspond to university lectures from two distinct faculties. Correspondingly, participants were sampled from these two faculties to ensure students from a particular faculty can recognize half of the lecture faces from their own faculty and have no knowledge for the rest half from the other one. The rest five were sourced from an online face database. In total, we recruited 21 participants. 

\subsection{Experimental Procedure}

\begin{figure}
    \centering
    \includegraphics[width=0.9\linewidth]{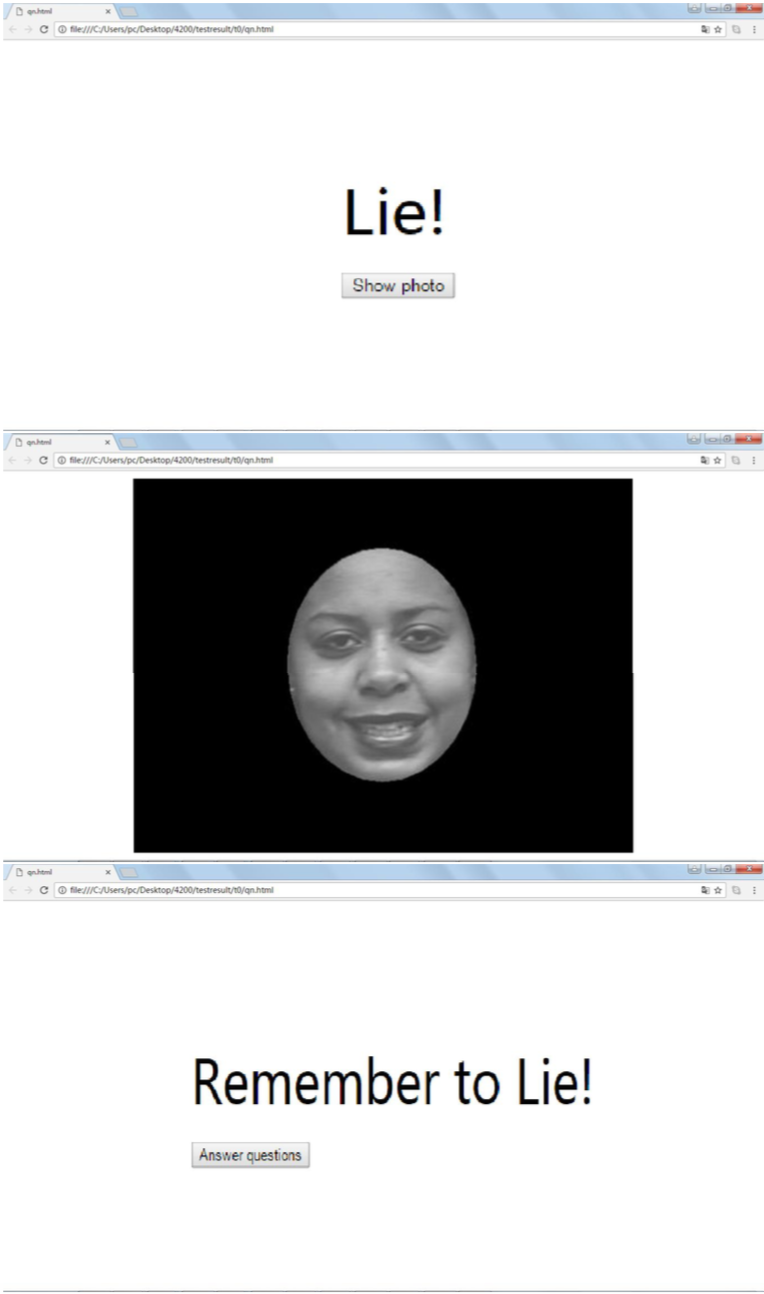}
    \caption{Screenshots of our user interface. }
    \label{fig:ui_prompt}
\end{figure}

We developed a web-browser based user interface to display our collected face images to the participants. During the experimental procedure, a participant will firstly see a huge instruction indicating either to ``tell the truth!'' or ``lie!'' for the upcoming face image. For example, if the face to display is familiar to the participant, he or she is expected to pretend to have no knowledge about the shown image. The same instruction will be prompted again to the participant before he or she gives the answer in order to avoid the unexpected honesty. We record participants' eye fixations trajectories during this process. Figure \ref{fig:ui_prompt} gives some screenshots of our user interface. 

\section{Results}
\label{sec:result}

The average fixation duration of the participants was 263.16 ms (SD = 49.34 ms). That is, the average number of fixation points in one trial of the experiment is 19. We analyzed a general eye movement pattern that summarizes all the 84 individual HMMs and four complete experimental settings of different combinations face familiarities and information concealing degrees, namely
\begin{enumerate}
    \item Telling the truth given a familiar face 
    \item Telling the truth given an unfamiliar face 
    \item Lying given a familiar face
    \item Lying given an unfamiliar face
\end{enumerate}
We next give the detailed eye movement patterns under different experimental settings, and will show that eye gaze trajectories and ROIs differ given different cognitive conditions during face recognition, namely different degrees of face familiarity and information concealing minds. To be more intuitive, all the left and right directions below are from the viewer's perspective. 

\subsection{General Eye Movement Pattern}

\begin{figure}
    \centering
    \includegraphics[width=0.9\linewidth]{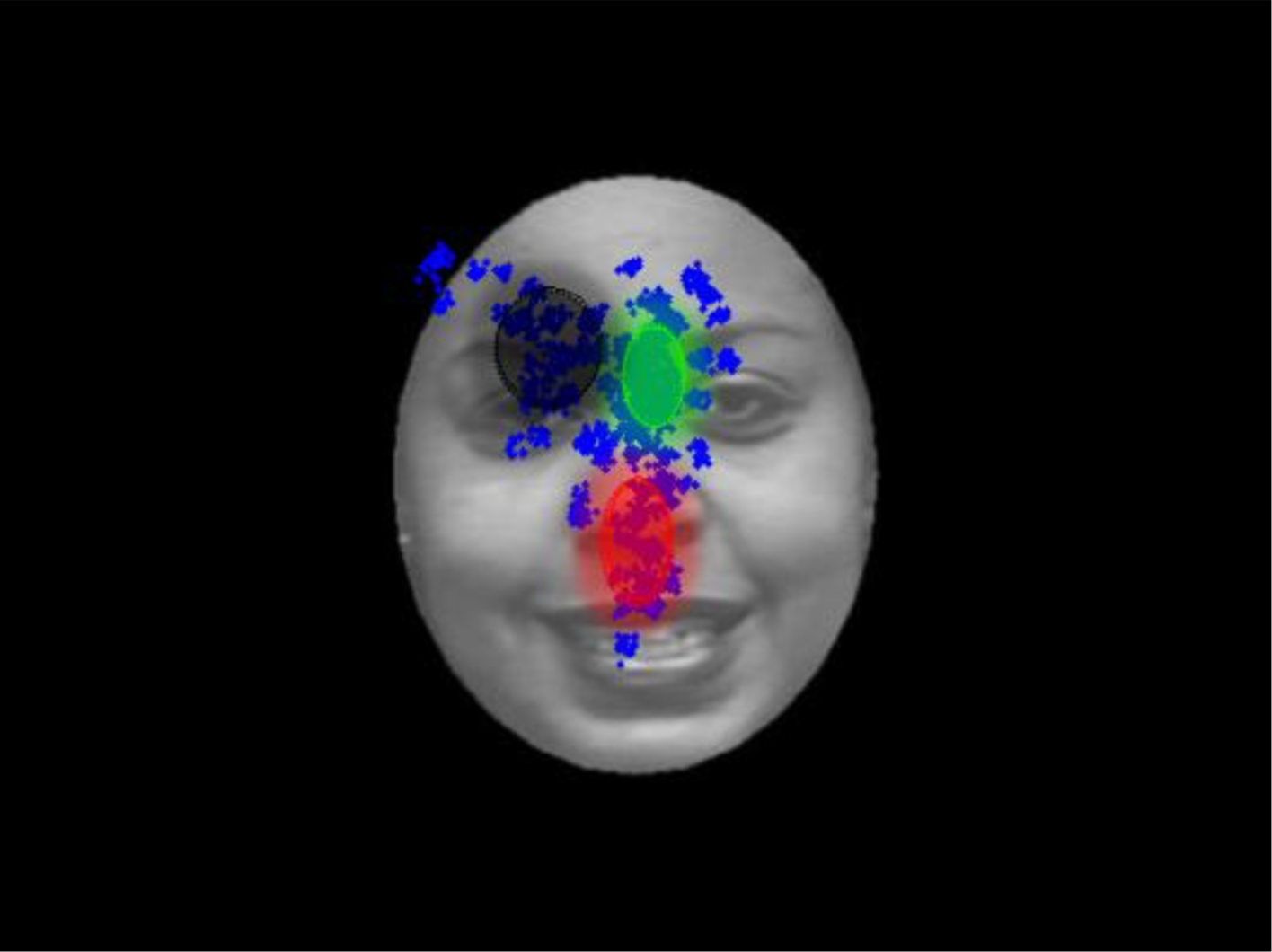}
    \caption{The general representative HMM that summarizes all the 84 individual HMMs. Blue dots are fixation points. Different colors represent distinct ROIs. }
    \label{fig:general_eye_movement_pattern}
\end{figure}

Following the methods given in \ref{subsec:eye_movement_analysis}, we obtained 84 Gaussian-HMMs, each with three means and covariance matrices to parametrize its three ROIs. We utilized the VHEM to group these 84 HMMs into a single representation that summarizes all the individual HMMs.
Figure \ref{fig:general_eye_movement_pattern} and Table \ref{tab:general_eye_movement_pattern} show the representative HMM that summarizes all the 84 participants' HMMs. 

\begin{table}[]
    \centering
    \begin{tabu} to \linewidth {X[3]X[2]X[2]X[2]} 
        \toprule
         Prior values & Red & Green & Blue \\
         \midrule
         & 0.0000 & 0.8475 & 0.1525 \\ 
         \toprule
         Transition probabilities & To Red & To Green& To Black \\ 
         \midrule
         From Red & 0.9704 & 0.0086 & 0.0210 \\
         From Green & 0.0449 & 0.9248 & 0.0303 \\
         From Blue & 0.0411 & 0.0248 & 0.9340 \\
         \bottomrule
    \end{tabu}
    \caption{Transition probabilities of the general representative HMM that summarizes all the 84 individual HMMs.}
    \label{tab:general_eye_movement_pattern}
\end{table}

From Figure \ref{fig:general_eye_movement_pattern} and Table \ref{tab:general_eye_movement_pattern}, the general scan path during this face recognition mostly likely starts from the green ROI (an oval region close to the inner canthus of the right eye), then goes to the red region (an oval region covers tip of the nose and the philtrum). Afterwards, the next fixation tends to remain in the red region. Also, the path is possible to move to the black region (close to the left eye, which is an oval region slightly above the eye and covers part of the left eyebrow). Finally, the probability of returning to the initial region (i.e. the green region) is quite low. In contrast, if starting the scan path from the black region with a low probability, it is highly possible that the path remains in the black region. 

\subsection{Telling the Truth Given a Familiar Face}

\begin{figure}
    \centering
    \includegraphics[width=0.9\linewidth]{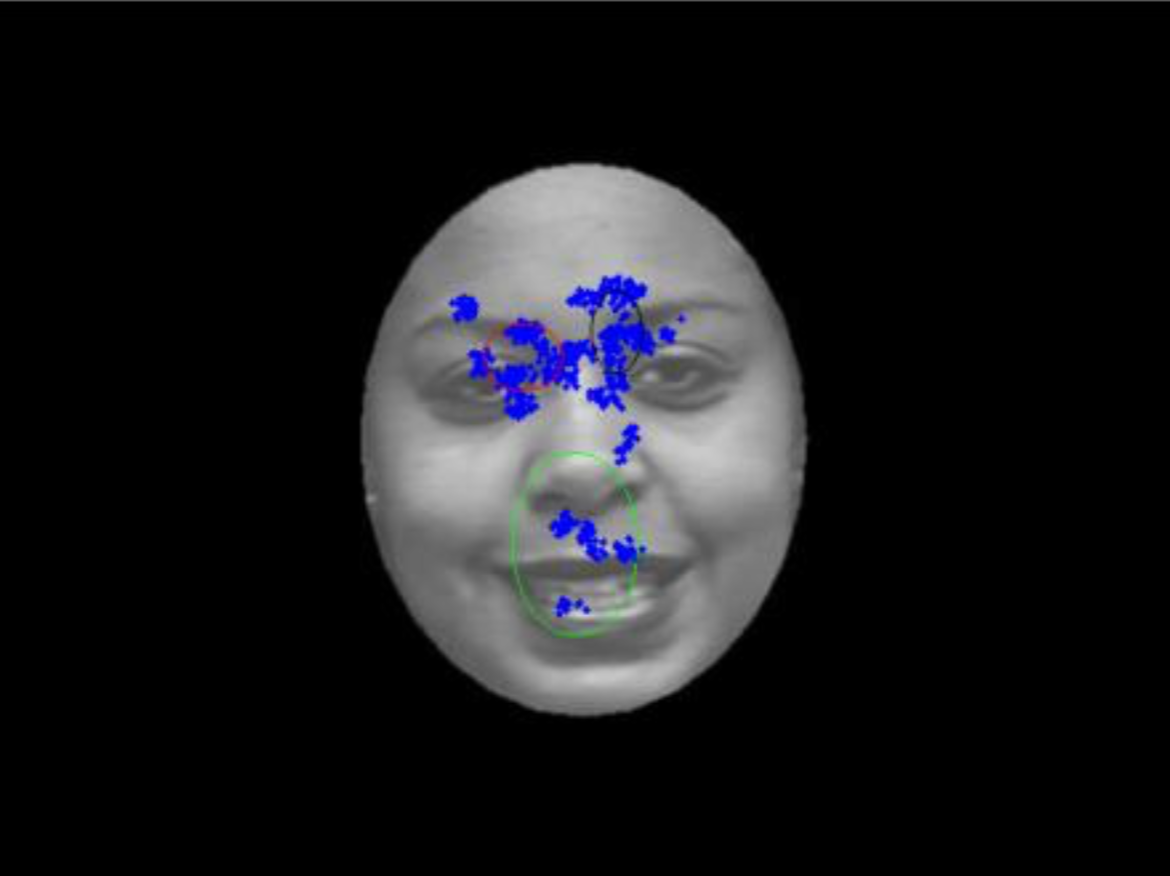}
    \caption{The ROIs for the eye movement pattern during telling the truth given a familiar face. }
    \label{fig:general_roi_true_familiar}
\end{figure}

\begin{figure}
    \centering
    \includegraphics[width=\linewidth]{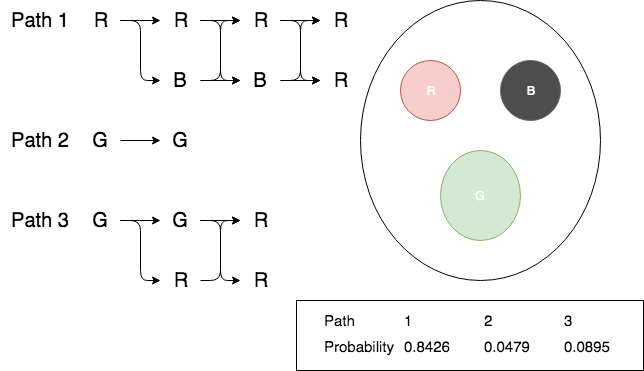}
    \caption{Scan paths for eye movement pattern during telling the truth given a familiar face. }
    \label{fig:scan_path_truth_familiar}
\end{figure}

As before, we applied VHEM and generated the HMM for the case of participants telling the truth given familiar faces. Figure \ref{fig:general_roi_true_familiar}, \ref{fig:scan_path_truth_familiar} and Table \ref{tab:general_roi_distribution_truth_familiar} illustrate more details on ROI distributions and scan moving paths. Specifically, the ROI distributions on the face of this case are: 
\begin{itemize}
    \item Red Region: Close to the inner canthus of the left eye from the viewer perspective, which is slightly above the eye, covering the inner part of the left eye and the left eyebrow, and the left upper eyelid. The semi-major axis is on the horizontal direction, while the semi-minor axis is on the vertical direction.
    
    \item Black Region: Close to the inner canthus of the right eye, which is slightly above the eye, covering a small inner part of the right eye, the right eyebrow and the right upper eyelid. The semi-major axis is on the vertical direction, while the semi-minor axis is on the horizontal direction.
    
    \item Green Region: The tip of the nose, the philtrum and most parts of the mouth. The mean axis of the oval region is slightly to the left compared to the middle axis of the face. The semi-major axis is on the vertical direction, while the semi-minor axis is the on horizontal direction.
\end{itemize}

In this sub-group, the first fixation tends to be in the red region (close to the left), followed by a high probability to remain in the red region, or it can move to the black region. Then it is highly probable to stay in the black region, if not, it can either move back and forth between the initial red region, or it can move to the green region, although the probability of moving to the green region is quite low. If the scan path starts from the black region by a low chance, the next fixation can possibly stay in the same black region, or it can move back and forth between the red and black regions. In this case, the probability of moving to the green region is still low. If the eye movement starts from the green region by a low probability, it is very likely to stay in the same region. The generated Viterbi path starts from the red region, then the black region, and concludes in the black region. 

\begin{table}[]
    \centering
    \begin{tabu} to \linewidth {X[3]X[2]X[2]X[2]} 
        \toprule
         Prior values & Red & Green & Blue \\
         \midrule
         & 0.8626 & 0.0479 & 0.0895 \\ 
         \toprule
         Transition probabilities & To Red & To Green& To Black \\ 
         \midrule
         From Red & 0.8859 & 0.0402 & 0.0738 \\
         From Green & 0.0285 & 0.9444 & 0.0272 \\
         From Blue & 0.0903 & 0.0252 & 0.8845 \\
         \bottomrule
    \end{tabu}
    \caption{Transition probabilities of the ROIs for eye movement pattern during telling the truth given a familiar face. }
    \label{tab:general_roi_distribution_truth_familiar}
\end{table}

\subsection{Telling the Truth on an Unfamiliar Face}

We applied the same approach to generate HMMs from participants telling the truth in front of unfamiliar faces. In this case, the specific ROI distributions on the face are: 

\begin{itemize}
    \item Red Region: A small part of the nose tip, a small part of the philtrum, and the a small part of the upper lip in the middle. The mean axis of this oval region is generally at the middle axis of the face. The semi-major axis is on vertical direction, while the semi-minor axis is on horizontal direction.
    
    \item Black Region: Close to the inner canthus of the right eye, which is being in a more inferior position to the right inner canthus, covering a small region close to the right inner canthus. The semi-major axis is on the vertical direction, while the semi-minor axis is on the horizontal direction.
    
    \item Green Region: A right part of the left eye, which is slightly to the central axis of the face, covering the most inner part of the left eye, the left eyebrow, and the region close to the left eye including a part of the nose bridge. The region is an approximate circle.
\end{itemize}

The most probable scan path starts from the black region (close to the right eye), with a high probability of remaining in the black region, or it can move to the red region or the green region (very unlikely). If it moves to the red region, next most likely action tends to stay in the same ROI, or it has a low probability to move to the green region. Otherwise, if it moves to the green region, it will likely to remain in the same region, to move to the red region by low chance. The probability of moving back to the black region is very low. In addition, it can start from green region, followed by a high probability of remaining in the green region. The probability of starting from the red region is very low, compared with from the other two ROIs. The Viterbi Path in this case implies that the most likely path is similar as the analysis above, of which starts from the the black region, then comes to the the red region, and end up with the the green region.

\begin{figure}
    \centering
    \includegraphics[width=0.9\linewidth]{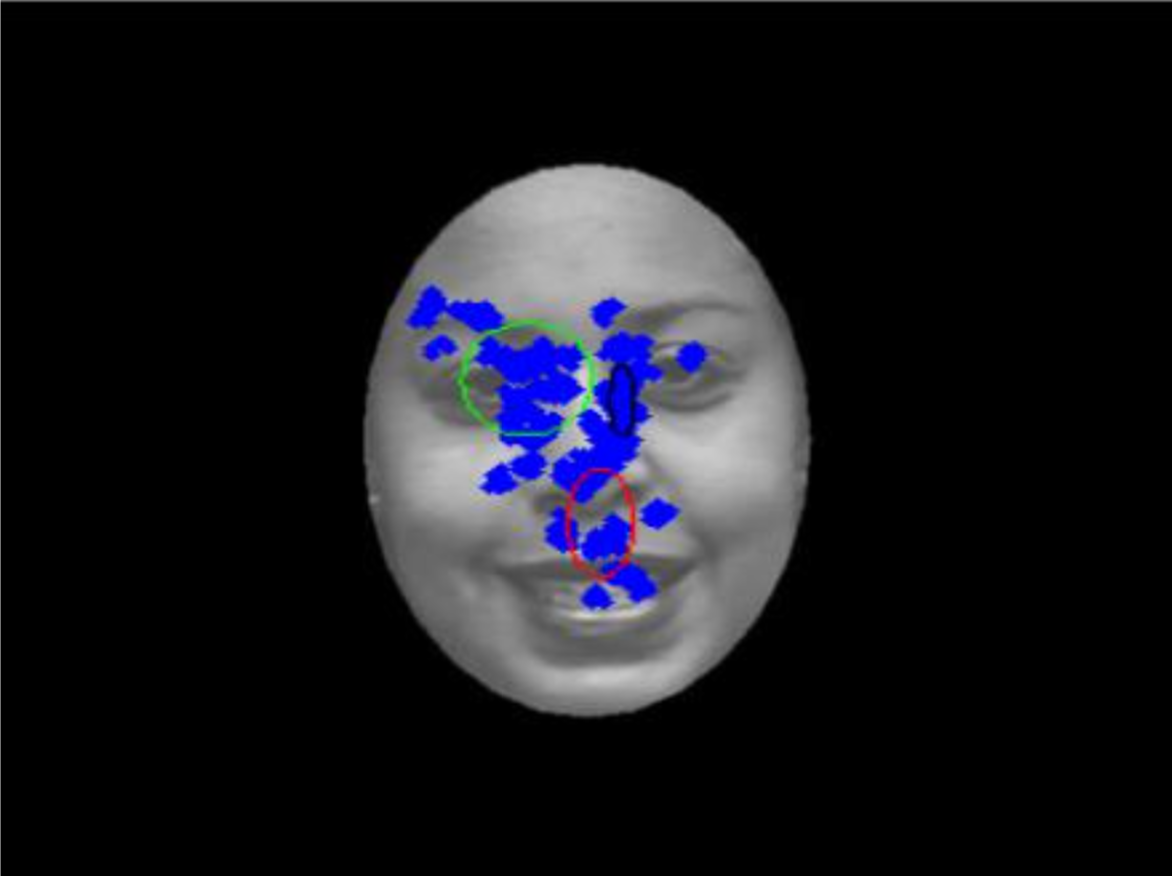}
    \caption{The ROIs for eye movement pattern during telling the truth given an unfamiliar face. }
    \label{fig:general_roi_truth_unfamiliar}
\end{figure}

\begin{table}[]
    \centering
    \begin{tabu} to \linewidth {X[3]X[2]X[2]X[2]} 
        \toprule
         Prior values & Red & Green & Blue \\
         \midrule
         & 0.0820 & 0.3354 & 0.5826 \\ 
         \toprule
         Transition probabilities & To Red & To Green& To Black \\ 
         \midrule
         From Red & 0.9680 & 0.0215 & 0.0105 \\
         From Green & 0.0518 & 0.9225 & 0.0257 \\
         From Blue & 0.0420 & 0.0898 & 0.8682 \\
         \bottomrule
    \end{tabu}
    \caption{Transition probabilities of the ROIs for eye movement pattern during telling the truth given an unfamiliar face. }
    \label{tab:general_roi_truth_unfamiliar}
\end{table}

\begin{figure}
    \centering
    \includegraphics[width=\linewidth]{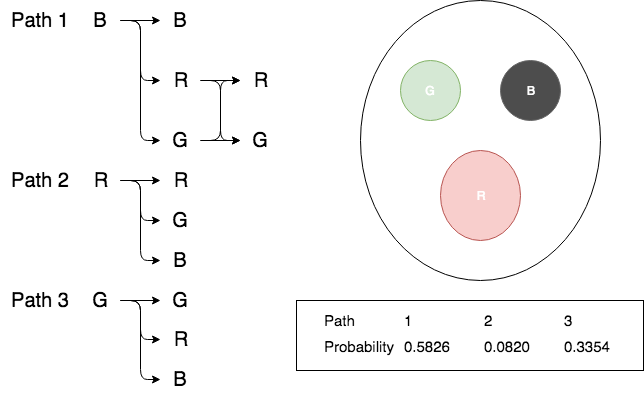}
    \caption{Scan paths for eye movement pattern during telling the truth given an unfamiliar face. }
    \label{fig:scan_path_truth_unfamiliar}
\end{figure}

\subsection{Lying Given a Familiar Face}

\begin{figure}
    \centering
    \includegraphics[width=0.9\linewidth]{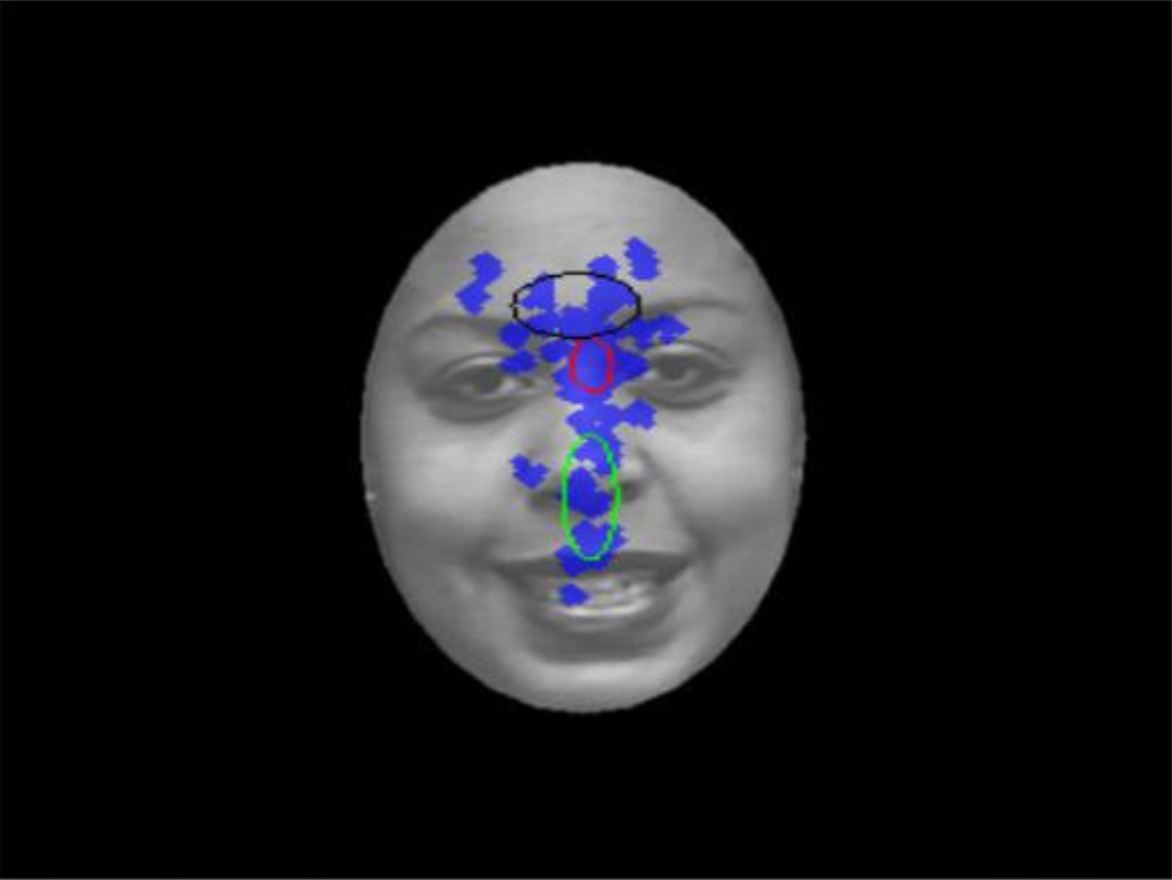}
    \caption{The ROIs for eye movement pattern during lying given a familiar face. }
    \label{fig:general_roi_lie_familiar}
\end{figure}

This situation shows a significant difference compared with the previous two cases. Figure \ref{fig:general_roi_lie_familiar} illustrates the representative Gaussian-HMM when participants lie about the familiarity while viewing the familiar face. As can be seen, the ROI distributions in face recognition of this case are: 
\begin{itemize}
    \item Red Region: A small part of the nose bridge between two eyes. The mean axis of this oval region is slightly to the right of the middle face axis. The oval region is an approximate circle.
    
    \item Black Region: The region between two eyebrows, covering part of the forehead close to the central of two eyebrows, which is slightly to the left of the middle axis of the face. The semi-major axis is on the horizontal direction, while the semi-minor axis is on the vertical direction.
    
    \item Green Region: Covering the lower part of the nose bridge, the tip of the nose, and the philtrum. The mean axis of this oval region is generally of the middle axis of the face. The semi-major axis is on the vertical direction, while the semi-minor axis is on the horizontal direction.
\end{itemize}

In this sub-group, the scan path most likely starts from the red region. Afterwards, it is more likely to remain in the same ROI, or with a lower probability, to move to the green or black regions, and stay the same locations subsequently. It is also possible to start from the black region, then with a higher likelihood of remaining the same region, or move to the other two regions. This Viterbi Path generally starts from the red region, then moves to the green region and converges to it.

\begin{table}[]
    \centering
    \begin{tabu} to \linewidth {X[3]X[2]X[2]X[2]} 
        \toprule
         Prior values & Red & Green & Blue \\
         \midrule
         & 0.6456 & 0.0000 & 0.3544 \\ 
         \toprule
         Transition probabilities & To Red & To Green& To Black \\ 
         \midrule
         From Red & 0.9102 & 0.0500 & 0.0398 \\
         From Green & 0.0119 & 0.9663 & 0.0218 \\
         From Blue & 0.0305 & 0.0501 & 0.9194 \\
         \bottomrule
    \end{tabu}
    \caption{Transition probabilities of the ROIs for eye movement pattern during lying given a familiar face. }
    \label{tab:general_roi_lie_familiar}
\end{table}

\begin{figure}
    \centering
    \includegraphics[width=\linewidth]{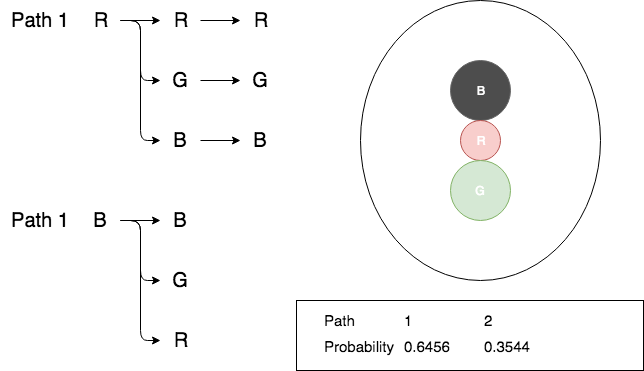}
    \caption{Scan paths for eye movement pattern during lying given a familiar face. }
    \label{fig:scan_path_lie_familiar}
\end{figure}

\subsection{Lying Given an Unfamiliar Face}

\begin{figure}
    \centering
    \includegraphics[width=\linewidth]{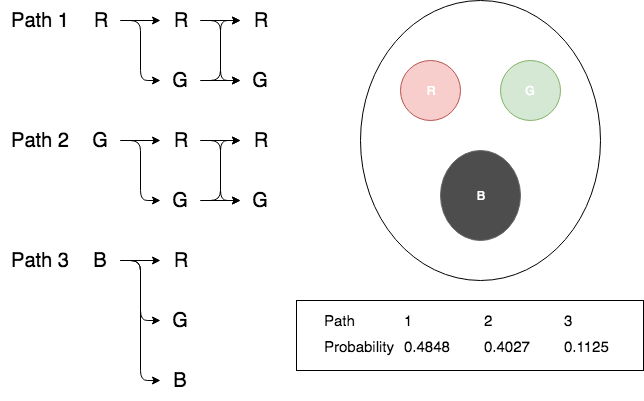}
    \caption{Scan paths for eye movement pattern during lying given an unfamiliar face. }
    \label{fig:scan_path_lie_unfamiliar}
\end{figure}

This case also displays a quite different eye behaviour compared with the previous ones. As Figure \ref{fig:scan_path_lie_unfamiliar} illustrates, the eye fixation points mainly locate on two eye regions, unlikely previously that there are a large proportion of points distribute to the nose region. The detailed ROI distributions are: 

\begin{itemize}
    \item Red Region: A small area located at the top-right left eye, slightly touching the left eyebrow, which is an approximate circle.
    
    \item Black Region: Covering a large area in the middle of the face including the nose, the philtrum, the lips, and the proximity regions. However, the fixation points are relatively sparse compared with the other two ROIs. 
    
    \item Green Region: A approximate circle shape distributed at the left part of the right eye, below the right eyebrow. 
\end{itemize}

In this case, it is equally likely for eye fixation points to locate in the red or green regions. If starting from the red region, the scan path is highly likely to remain in the same region, or move to the green region, move back and forth. Moving to the black region is relatively unlikely to happen. The Viterbi path shows that this scan path is probable to conclude with an oscillation between the red and green regions. 

\begin{figure}
    \centering
    \includegraphics[width=0.9\linewidth]{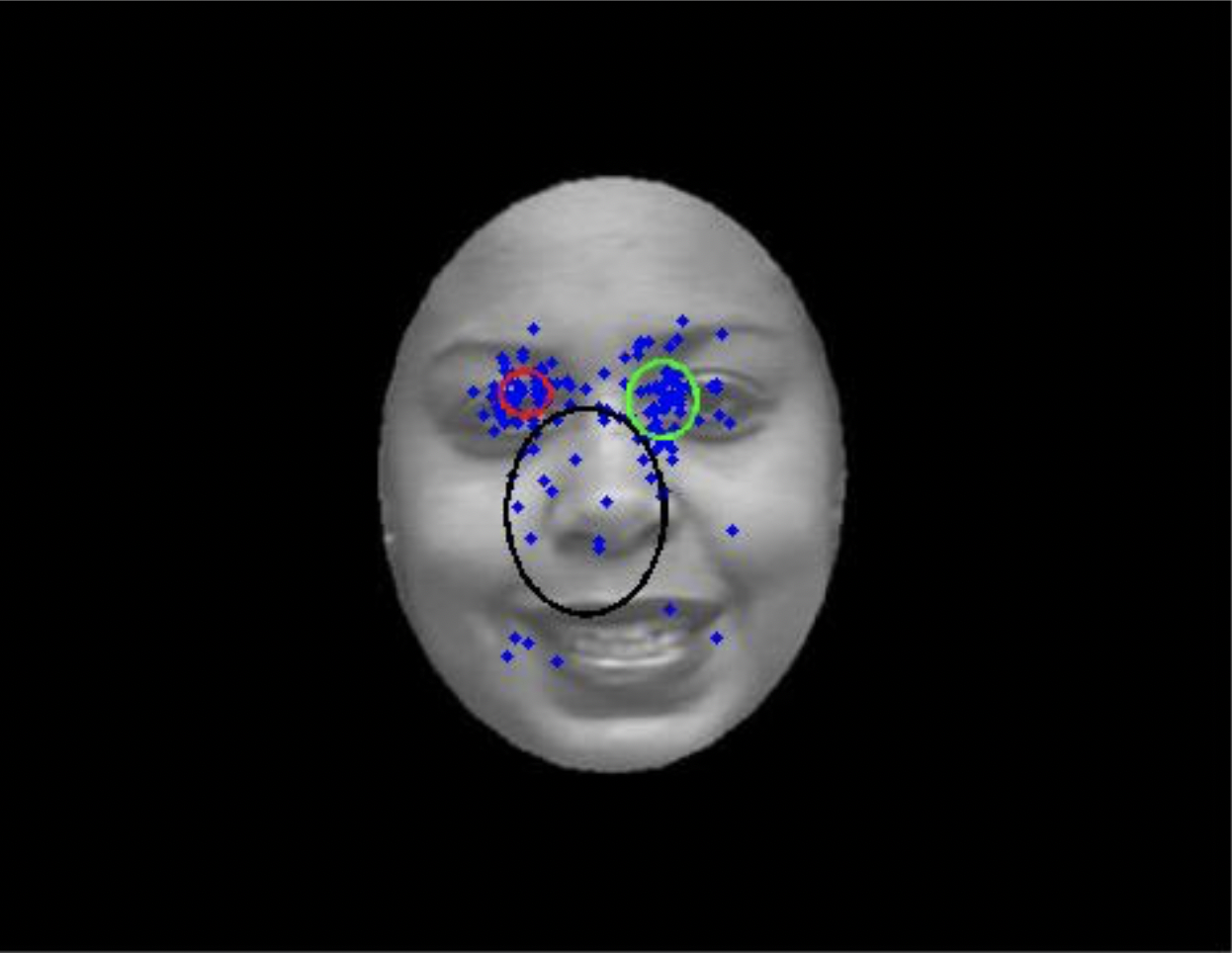}
    \caption{The ROIs for eye movement pattern during lying given an unfamiliar face. }
    \label{fig:general_roi_lie_unfamiliar}
\end{figure}

\begin{table}[]
    \centering
    \begin{tabu} to \linewidth {X[3]X[2]X[2]X[2]} 
        \toprule
         Prior values & Red & Green & Blue \\
         \midrule
         & 0.4848 & 0.4027 & 0.1125 \\ 
         \toprule
         Transition probabilities & To Red & To Green& To Black \\
         \midrule
         From Red & 0.5602 & 0.3775 & 0.0623 \\
         From Green & 0.3153 & 0.5622 & 0.1225 \\
         From Blue & 0.2099 & 0.2751 & 0.5150 \\
         \bottomrule
    \end{tabu}
    \caption{Transition probabilities of the ROIs for eye movement pattern during lying given an unfamiliar face. }
    \label{tab:general_roi_lie_unfamiliar}
\end{table}

\subsection{Model Generalization Test}

In this section, we show our tested results for the generalization of our proposed eye movement patterns for different face familiarity and information concealing. We generated a Viterbi path through the HMM methodology for each of our 21 participants. This is followed by calculating four Euclidean distances between this individual Viterbi path and the general Viterbi paths corresponding the previous four conditions, namely truth telling or lying in front of familiar and unfamiliar faces. 

We evaluated the accuracy results of classifying eye movement patterns into telling the truth or a lie for both familiar and unfamiliar face recognition. The outcomes are well-beyond the chance, as Tables \ref{tab:familiar_truth_classification} and \ref{tab:unfamiliar_truth_classification} demonstrate. Furthermore, we also classify it to one of the four conditions according to the shortest Euclidean path. The 1-in-4 classifying accuracies are in Table \ref{tab:four_classification}. Again, as this table shows, eye movement patterns are more distinguishable when participants are lying. These results reveal a promising possibility of utilizing eye movement patterns to analyze latent cognitive activities. 

\begin{table}[]
    \centering
    \begin{tabu}to \linewidth {X[2]X[2]X[2]X[2]X[2]} 
    \toprule
    & Familiar \& Truth & Unfamiliar \& Truth & Familiar \& Lie & Unfamiliar \& Lie \\ 
    \midrule
         Accuracy & 66.67\% (14/21) & 61.90\% (13/21) & 76.19\% (13/21) & 71.43\% (15/21) \\
    \bottomrule
    \end{tabu}
    \caption{Accuracy results of classifying eye movement patterns to their source face familiarity and degrees of information concealing. }
    \label{tab:four_classification}
\end{table}

\begin{table}[]
    \centering
    \begin{tabu}to \linewidth {X[2]X[4]X[4]} 
    \toprule
    & Familiar \& Truth & Familiar \& Lie \\ 
    \midrule
         Accuracy & 80.95\% (17/21) & 85.71\% (18/21) \\
    \bottomrule
    \end{tabu}
    \caption{Accuracy results of classifying eye movement patterns of familiar face recognition when telling the truth or a lie. }
    \label{tab:familiar_truth_classification}
\end{table}

\begin{table}[]
    \centering
    \begin{tabu}to \linewidth {X[2]X[4]X[4]} 
    \toprule
    & Familiar \& Truth & Familiar \& Lie \\ 
    \midrule
         Accuracy & 71.43\% (15/21) & 80.95\% (17/21) \\
    \bottomrule
    \end{tabu}
    \caption{Accuracy results of classifying eye movement patterns of unfamiliar face recognition when telling the truth or a lie. }
    \label{tab:unfamiliar_truth_classification}
\end{table}
\section{Conclusion}

In this study, we generalize five representative eye movement patterns, characterized by their trajectories and fixation point distributions. These five patterns include telling the truth or a lie in front of a familiar or unfamiliar face, and a general one using all the individual data. We found that the general eye movement patterns of lying, during both a familiar and unfamiliar face recognition, are significantly different comparing with truth-telling situations. Subsequent tests using the four eye movement patterns generated using HMMs with Gaussian emission, except the general one, demonstrated a good performance for discerning deception.

In the future, we will compose our eye movement patterns into deception detection systems as additional channel to see whether they can help enhancing the original performance of these systems. Moreover, we will test the feasibility of whether eye movement patterns behave differently under different cognitive modes such as happiness and anger. 

{\small
\bibliographystyle{ieee}
\bibliography{egbib}
}

\end{document}